\makeatletter
\newcommand{\dontusepackage}[2][]{%
  \@namedef{ver@#2.sty}{9999/12/31}%
  \@namedef{opt@#2.sty}{#1}}
\makeatother
\dontusepackage{subfigure}

\documentclass[]{article}

\usepackage{lmodern}
\usepackage{algorithm}
\usepackage{algpseudocode}

\usepackage{amssymb,amsmath}
\usepackage{ifxetex,ifluatex}
\usepackage[usenames,dvipsnames]{color}
\usepackage{fixltx2e} 
\ifnum 0\ifxetex 1\fi\ifluatex 1\fi=0 
  \usepackage[T1]{fontenc}
  \usepackage[utf8]{inputenc}
\else 
  \ifxetex
    \usepackage{mathspec}
    \usepackage{xltxtra,xunicode}
  \else
    \usepackage{fontspec}
  \fi
  \defaultfontfeatures{Mapping=tex-text,Scale=MatchLowercase}
  
\fi
\IfFileExists{upquote.sty}{\usepackage{upquote}}{}
\IfFileExists{microtype.sty}{%
\usepackage{microtype}
\UseMicrotypeSet[protrusion]{basicmath} 
}{}
\usepackage[margin=1.0in,bottom=1.5in]{geometry}
\usepackage[square,numbers,sort&compress]{natbib}
\usepackage{listings}
\usepackage{graphicx}
\makeatletter
\def\maxwidth{\ifdim\Gin@nat@width>\linewidth\linewidth\else\Gin@nat@width\fi}
\def\maxheight{\ifdim\Gin@nat@height>\textheight\textheight\else\Gin@nat@height\fi}
\makeatother
\setkeys{Gin}{width=\maxwidth,height=\maxheight,keepaspectratio}
\usepackage{caption}
\usepackage{float}
\usepackage{subfig}
\usepackage{seqsplit}
\ifxetex
  \usepackage[setpagesize=false, 
              unicode=false, 
              xetex]{hyperref}
\else
  \usepackage[unicode=true]{hyperref}
\fi
\hypersetup{breaklinks=true,
            bookmarks=true,
            pdfauthor={},
            pdftitle={WISER: multimodal variational inference for full-waveform inversion without dimensionality reduction},
            colorlinks=true,
            citecolor=black,
            urlcolor=blue,
            linkcolor=black,
            pdfborder={0 0 0}}
\usepackage[preprint]{neurips_2019}
\usepackage{url}
\usepackage{booktabs}
\usepackage{amsfonts}
\usepackage{nicefrac}
\usepackage{microtype}

\makeatletter
\let\@oldfnsymbol\@fnsymbol
\renewcommand{\@fnsymbol}[1]{\@oldfnsymbol{0}}
\makeatother

\title{WISER: multimodal variational inference for full-waveform inversion without dimensionality reduction}
\author{
    Ziyi Yin\textsuperscript{1}\thanks{$^1$Correspondence: \href{mailto:ziyi.yin@gatech.edu}{ziyi.yin@gatech.edu}},
    Rafael Orozco\textsuperscript{1},
    Felix J. Herrmann\textsuperscript{1}\\
    \textsuperscript{1}Georgia Institute of Technology
}

\begin{document}
\maketitle

\begin{abstract}
We present a semi-amortized variational inference framework designed for
computationally feasible uncertainty quantification in 2D full-waveform
inversion to explore the multimodal posterior distribution without
dimensionality reduction. The framework is called \textbf{WISER}, short
for full-\textbf{W}aveform variational \textbf{I}nference via
\textbf{S}ubsurface \textbf{E}xtensions with \textbf{R}efinements. WISER
leverages the power of generative artificial intelligence to perform
approximate amortized inference that is low-cost albeit showing an
amortization gap. This gap is closed through non-amortized refinements
that make frugal use of acoustic wave physics. Case studies illustrate
that WISER is capable of full-resolution, computationally feasible, and
reliable uncertainty estimates of velocity models and imaged
reflectivities.
\end{abstract}

\section{Introduction}\label{introduction}

Full-waveform inversion (FWI) aims to estimate unknown multi-dimensional
($\mathrm{D}\geq 2$) velocity models, denoted as $\mathbf{x}$, from
noisy seismic data, $\mathbf{y}$, by inverting the nonlinear forward
operator, $\mathcal{F}$, which relates $\mathbf{x}$ and $\mathbf{y}$ via
$\mathbf{y} = \mathcal{F}(\mathbf{x})+\boldsymbol{\epsilon}$ with
$\boldsymbol{\epsilon}$ measurement noise \citep{virieux2009}. FWI poses
significant challenges, as it requires solving a high-dimensional,
non-convex, and ill-posed inverse problem, with a computationally
demanding forward operator in multiple dimensions. In addition, the
inherent nonuniqueness of FWI results leads to multiple possible Earth
models compatible with the observed data, underscoring the need for
uncertainty quantification (UQ) to handle this multimodality.

The trade-off between accuracy and computational cost is a critical
consideration in any high-dimensional inference routine with expensive
forward operators \citep{cranmer2020frontier}, including FWI. To
circumvent the costs associated with global optimization, several
approaches have attempted localized UQ
\citep{fang2018uncertainty, keating2021null, izzatullah2023physics, hoffmann2024local}
based on the Laplace approximation. However, these approaches may not
capture the full complexities of multimodal parameter spaces. In
contrast, a Bayesian inference approach offers a costly but
comprehensive resolution of the posterior distribution,
$p(\mathbf{x}\mid\mathbf{y})$.

Bayesian inference algorithms are broadly categorized into two groups.
The first, sampling-based methods, like Markov-chain Monte Carlo
\citep[McMC,][]{cowles1996markov}, struggle with high-dimensional
parameter spaces. To meet this challenge, they often rely on too
restrictive low-dimensional parameterizations to reduce the number of
sampling iterations
\citep{fang2020deep, liang2023uncertainty, wei2023quantifying, wei20243d, wei2024reconstruction, dhabaria2024hamiltonian},
which could bias the inference results, rendering them impractical for
multi-D UQ studies especially when solutions are nonunique.

The second category, optimization-based methods, like variational
inference \citep[VI,][]{zhang2021introduction}, seek to approximate the
posterior distribution using classes of known parameterized
distributions. VI can be subdivided into amortized and non-amortized
methods. Amortized VI involves a computationally intensive offline
training phase, leveraging advances in generative artificial
intelligence (genAI), particularly with models like conditional
diffusion \citep{baldassari2023conditional} and conditional normalizing
flows \citep[CNFs,][]{winkler2019learning}. After training, amortized VI
provides rapid sampling during inference
\citep{siahkoohi2023, orozco2023amortized}, exemplified by the WISE
framework \citep{yin2024wise} for multi-D FWI problems. However, these
methods may suffer from an \emph{amortization gap} --- implying that the
amortized networks may only deliver suboptimal inference for a single
observation at inference time, particularly when trained with limited
examples or when there exists a discrepancy between training and
inference \citep{marino2018iterative}. Conversely, non-amortized VI
dedicates entire computational resources to the online inference
\citep{zhao2022bayesian, zhang20233, zhang2023bayesian}. They result in
more accurate inference, but the costly optimization has to be carried
out repeatedly for new observations, and integrating realistic priors
can be challenging since the prior needs to be embedded involving
density evaluations \citep{kruse2021hint}.

This paper introduces WISER as a semi-amortized VI framework to
facilitate computationally feasible and reliable UQ for multi-D FWI
without dimensionality reduction. Building on WISE, we train CNFs for
efficient, suboptimal amortized inference, but then follow up with a
crucial refinement step that only needs frugal use of the forward
operator and its gradient. The refinement step aligns the posterior
samples with the observation during inference, effectively bridging the
amortization gap and enhancing inference accuracy.

Our contributions are organized as follows. We begin by outlining WISER
in Algorithm~\ref{alg-wiser}. We explore the algorithm by explaining amortized
VI with WISE, followed by computationally feasible multi-D physics-based
refinement. The performance of WISER is demonstrated through realistic
synthetic 2D case studies using the Compass model \citep{e.jones2012},
showcasing improvements over WISE for both in- and out-of-distribution
scenarios.

\begin{algorithm}[htb!]
\caption{WISER: full-\textbf{W}aveform variational \textbf{I}nference via \textbf{S}ubsurface \textbf{E}xtensions with \textbf{R}efinements}
\label{alg-wiser}
\begin{algorithmic}[1]
\State {\large\textbf{Offline training phase}}
\State
\State \textbf{Dataset generation}
\For{$i = 1:N$}
  \State $\mathbf{x}^{(i)} \sim p(\mathbf{x})$
  \State $\boldsymbol{\epsilon}^{(i)} \sim p(\boldsymbol{\epsilon})$
  \State $\mathbf{y}^{(i)} = \mathcal{F}(\mathbf{x}^{(i)})+\boldsymbol{\epsilon}^{(i)}$
  \State $\overline{\mathbf{y}}^{(i)} = \overline{\nabla\mathcal{F}}(\mathbf{x}_0)^{\top}\mathbf{y}^{(i)}$
\EndFor
\State
\State \textbf{Network training}
\State $\displaystyle \boldsymbol{\theta}^{\ast}=\underset{\boldsymbol{\theta}}{\operatorname{argmin}} \quad \frac{1}{N}\sum_{i=1}^{N} \left(\frac{1}{2}\|f_{\boldsymbol{\theta}}\left(\mathbf{x}^{(i)};\overline{\mathbf{y}}^{(i)}\right)\|_2^2-\log\left|\det\mathbf{J}_{f_{\boldsymbol{\theta}}}\right|\right)$
\State
\State {\large\textbf{Online inference phase}}
\State
\State $\overline{\mathbf{y}}_{\mathrm{obs}} = \overline{\nabla\mathcal{F}}(\mathbf{x}_0)^{\top}\mathbf{y}_{\mathrm{obs}}$
\For{$i = 1:M$}
\State $\mathbf{z}_i \sim \mathrm{N}(\mathbf{0}, \mathbf{I})$
\State $\mathbf{x}_i = f_{\boldsymbol{\theta}^{\ast}}^{-1}\left(h_{\boldsymbol{\phi}}\left(\mathbf{z}_i\right);\overline{\mathbf{y}}_{\mathrm{obs}}\right)$
\EndFor
\State
\State \textbf{Physics-based refinements}
\For{$ii = 1:\mathrm{maxiter}_1$}
  \For{$i=1:M$}
        \State $\displaystyle\mathbf{g}_i=\nabla_{\mathbf{x}_i}\left[\frac{1}{2\sigma^2}\|\mathcal{F}(\mathbf{x}_i)-\mathbf{y}_{\mathrm{obs}}\|_2^2 + \frac{1}{2\gamma^2}\|\mathbf{x}_i-f_{\boldsymbol{\theta}^{\ast}}^{-1}\left(h_{\boldsymbol{\phi}}\left(\mathbf{z}_i\right);\overline{\mathbf{y}}_{\mathrm{obs}}\right)\|_2^2\right]$
        \State $\mathbf{x}_i = \mathbf{x}_i - \tau \mathbf{g}_i$
  \EndFor
  \For{$iii= 1:\mathrm{maxiter}_2$}
  \State $\displaystyle\mathcal{L}(\boldsymbol{\phi})=\sum_{i=1}^M \left[\frac{1}{2\gamma^2}\|\mathbf{x}_i-f_{\boldsymbol{\theta}^{\ast}}^{-1}\left(h_{\boldsymbol{\phi}}\left(\mathbf{z}_i\right);\overline{\mathbf{y}}_{\mathrm{obs}}\right)\|_2^2 + \frac{1}{2}\|h_{\boldsymbol{\phi}}\left(\mathbf{z}_i\right)\|_2^2 - \log\left|\det\mathbf{J}_{h_{\boldsymbol{\phi}}}\right|\right]$
  \State $\boldsymbol{\phi} \leftarrow \mathrm{ADAM}\left(\mathcal{L}(\boldsymbol{\phi})\right)$
  \EndFor
\EndFor
\State
\State {\large\textbf{Output}}: $\{f_{\boldsymbol{\theta}^{\ast}}^{-1}\left(h_{\boldsymbol{\phi}}\left(\mathbf{z}_i\right);\overline{\mathbf{y}}_{\mathrm{obs}}\right)\}_{i=1}^M$ as samples of $p(\mathbf{x}|\mathbf{y}_{\mathrm{obs}})$
\end{algorithmic}
\end{algorithm}

\section{Amortized VI with WISE (lines
1---20)}\label{amortized-vi-with-wise-lines-120}

WISER starts with an offline training phase that leverages conditional
generative models to approximate the posterior distribution. This is
achieved by WISE \citep{yin2024wise}, which involves generating a
training dataset (lines 3---9 of Algorithm~\ref{alg-wiser}) and training the
CNFs (line 11---12).

\subsection{Dataset generation (lines
3---9)}\label{dataset-generation-lines-39}

We begin by drawing $N$ velocity models from the prior distribution,
denoted by $p(\mathbf{x})$ (line 5). For each sample,
$\mathbf{x}^{(i)}$, we simulate the observed data, $\mathbf{y}^{(i)}$,
by performing the wave modeling and adding a random noise term (lines
6---7). Next, we compute common-image gathers \citep[CIGs,][]{hou2016}
for each observed data with an initial smooth 1D migration-velocity
model, $\mathbf{x}_0$, which can be rather inaccurate. These CIGs,
represented by $\overline{\mathbf{y}}^{(i)}$, are produced by applying
the adjoint of the extended migration operator,
$\overline{\nabla\mathcal{F}}(\mathbf{x}_0)^{\top}$, to the observed
data. Using CIGs as the set of physics-informed summary statistics not
only preserves information from the observed seismic data
\citep{ten2023omnidirectional} but also enhances the training of CNFs in
the next stage \citep{radev2020bayesflow, orozco2023}, as they help to
decode the wave physics, translating prestack data to the image
(subsurface-offset) domain.

\subsection{Network training (lines
11---12)}\label{network-training-lines-1112}

CNFs are trained with pairs of velocity models and CIGs via minimization
of the objective in line 12. The symbol $f_{\boldsymbol{\theta}}$
denotes the CNFs, characterized by their network weights,
$\boldsymbol{\theta}$, and the Jacobian,
$\mathbf{J}_{f_{\boldsymbol{\theta}}}$. The term ``normalizing'' within
CNFs implies their functionality to transform realizations of velocity
models, $\mathbf{x}^{(i)}$, into Gaussian noise from a standard
multivariate normal distribution (as defined by the $\ell_2$ norm),
conditioned on the summary statistics (CIGs).

\subsection{Online inference (lines
14---20)}\label{online-inference-lines-1420}

The aforementioned data generation and CNF training procedures conclude
the offline training phase. During online inference, amortized VI is
enabled by leveraging the inherent invertibility of CNFs. For a given
observation, $\mathbf{y}_{\mathrm{obs}}$, the online cost is merely
generation of a single set of CIGs (line 16). Subsequently, the
posterior samples are generated by applying the inverse of the CNFs to
Gaussian noise realizations, conditioned on these CIGs (lines
18---19)\footnote{We slightly abuse the notation to assume
  $h_{\boldsymbol{\phi}}$ as an identity operator here.}.

\section{Physics-based refinements (lines
22---32)}\label{physics-based-refinements-lines-2232}

Consider a single observation, $\mathbf{y}_{\mathrm{obs}}$, and its
corresponding posterior samples,
$\mathbf{x}_i \sim p(\mathbf{x} \mid \overline{\mathbf{y}}_{\mathrm{obs}})$.
The latent representations generated by the trained CNFs,
${\hat{\mathbf{z}}_i} = f_{\boldsymbol{\theta}^{\ast}}(\mathbf{x}_i; \overline{\mathbf{y}}_{\mathrm{obs}})$,
may not conform exactly to the standard Gaussian distribution during
inference. To address this issue, we follow \citet{siahkoohi2023} to
apply latent space corrections to fine-tune the trained CNFs. This
involves integrating a shallower, yet invertible, network\footnote{For
  linear inverse problems in seismic imaging, \citet{siahkoohi2023} show
  that an elementwise scaling and shift mechanism is adequate to bridge
  the gap. However, given the complex, non-convex nature of FWI, we
  employ $h_{\boldsymbol{\phi}}$ as a generic invertible network.},
specifically trained to map realizations of true Gaussian noise to the
corresponding latent codes, ${\hat{\mathbf{z}}_i}$. Adhering to a
transfer learning approach, we maintain the weights of the trained CNFs
while solely updating the weights of the shallower network by minimizing
the following objective:

\begin{equation}
\underset{\boldsymbol{\phi}}{\operatorname{minimize}} \quad \mathbb{E}_{\mathbf{z}\sim\mathrm{N}(\mathbf{0}, \mathbf{I})}\left[\frac{1}{2\sigma^2}\|\mathcal{F}\circ f_{\boldsymbol{\theta}^{\ast}}^{-1}\left(h_{\boldsymbol{\phi}}\left(\mathbf{z}\right);\overline{\mathbf{y}}_{\mathrm{obs}}\right)-\mathbf{y}_{\mathrm{obs}}\|_2^2  + \frac{1}{2}\|h_{\boldsymbol{\phi}}\left(\mathbf{z}\right)\|_2^2- \log\left|\det\mathbf{J}_{h_{\boldsymbol{\phi}}}\right|\right].
\label{eq-strong}
\end{equation}

Here, the refinement network, $h_{\boldsymbol{\phi}}$, mitigates the
amortization gap by adjusting the latent variable $\mathbf{z}$ before
feeding it to the inverse of the trained CNFs,
$f_{\boldsymbol{\theta}^\ast}^{-1}$. Intuitively, minimizing the first
terms ties the posterior samples closer to the observed data. The second
and third terms prevent the corrected latent space from being far from
the Gaussian distribution, which implicitly takes advantage of the prior
information existing in the amortized training phase.

Equation~\ref{eq-strong} offers a fine-tuning approach that leverages the full
multi-D wave physics to refine the amortized VI framework for a single
observation at inference phase. However, it introduces notable
computational demands because it necessitates the coupling of the
modeling operator and the networks. Specifically, every update to the
network weights, $\boldsymbol{\phi}$, requires costly wave modeling
operations. Given that network training typically involves numerous
iterations, these computational demands can render it impractical for
realistic FWI applications.

To relieve this computational burden, we adopt a strategy from
\citet{siahkoohi2020weak} to reformulate Equation~\ref{eq-strong} into a weak
form by allowing the network output to be only weakly enforced (in an
$\ell_2$ sense) to be the corrected velocity models. The objective
function for this weak formulation reads:

\begin{equation}
 \underset{\mathbf{x}_{1:M}, \boldsymbol{\phi}}{\operatorname{minimize}} \quad \frac{1}{M}\sum_{i=1}^{M} \left[\frac{1}{2\sigma^2}\|\mathcal{F}(\mathbf{x}_i)-\mathbf{y}_{\mathrm{obs}}\|_2^2
+ \frac{1}{2\gamma^2}\|\mathbf{x}_i-f_{\boldsymbol{\theta}^{\ast}}^{-1}\left(h_{\boldsymbol{\phi}}\left(\mathbf{z}_i\right);\overline{\mathbf{y}}_{\mathrm{obs}}\right)\|_2^2 + \frac{1}{2}\|h_{\boldsymbol{\phi}}\left(\mathbf{z}_i\right)\|_2^2 - \log\left|\det\mathbf{J}_{h_{\boldsymbol{\phi}}}\right| \right].
\label{eq-weak}
\end{equation}

We strategically decouple the computationally expensive forward
operator, $\mathcal{F}$, from the more cheap-to-evaluate networks,
$f_{\boldsymbol{\theta}^\ast}$ and $h_{\boldsymbol{\phi}}$. This is
achieved in a penalty form with the assumption that the misfit between
the network outputs and the posterior samples adheres to a Gaussian
distribution, $\mathrm{N}(\mathbf{0}, \gamma^2\mathbf{I})$, where
$\gamma$ is a hyperparameter dictating the trade-off between data misfit
and regularization. Setting $\gamma$ to $0$ equates this weak
formulation to the constrained formulation in Equation~\ref{eq-strong}. This
weak formulation also supports optimization strategies for updating the
velocity models with physical constraints
\citep{esser2018total, peters2019projection} and multiscale optimization
techniques \citep{bunks1995multiscale}.

WISER takes full computational advantage of this weak formulation by
employing a nested loop structure. The outer loop is dedicated to
updating $M$ velocity models, $\mathbf{x}_i$, through costly gradient
descent steps (lines 24---27 of Algorithm~\ref{alg-wiser}), while the inner loop
(lines 28---31) focuses on more updates (with the ADAM optimizer) to
network weights, $\boldsymbol{\phi}$, without computationally expensive
physics modeling. To achieve a balance, we conduct
$\mathrm{maxiter}_2=128$ iterations in the inner loop. After
refinements, WISER first evaluates the refined network on the latent
variables to obtain refined latent codes. Subsequently, the amortized
network uses the refined codes conditioned on the CIGs to compute the
corrected posterior samples (line 34).

\section{Case studies}\label{case-studies}

Evaluation of WISER is conducted through synthetic case studies
utilizing 2D slices of the Compass model and 2D acoustic wave physics.
The parameter of interest is discretized into $512 \times 256$
gridpoints with a spatial resolution of $12.5\,\mathrm{m}$, resulting in
over $10^5$ degrees of freedom. The forward operator, $\mathcal{F}$,
simulates acoustic data with absorbing boundaries. A Ricker wavelet with
a central frequency of $15\,\mathrm{Hz}$ and an energy cut below
$3\,\mathrm{Hz}$ is employed. We use $512$ sources towed at
$12.5\,\mathrm{m}$ depth and $64$ ocean-bottom nodes (OBNs) located at
jittered sampled horizontal positions \citep{hennenfent2008simply}. We
employ source-receiver reciprocity during the modeling and sensitivity
calculations. The observed data, $\mathbf{y}^{\mathrm{obs}}$, is
perturbed with band-limited Gaussian noise to achieve a signal-to-noise
ratio (S/N) of $12\,\mathrm{dB}$. The training of the CNFs uses $N=800$
pairs of velocity models and CIGs. To demonstrate WISER's efficacy in
mitigating the amortization gap, we compare results from WISE and WISER
under two scenarios during inference:

\begin{enumerate}
\def\labelenumi{(\roman{enumi})}
\itemsep1pt\parskip0pt\parsep0pt
\item
  observed shot data is generated using an in-distribution velocity
  model with the same forward operator;
\item
  observed shot data is produced by an out-of-distribution (OOD)
  velocity model and also a slightly altered forward operator.
\end{enumerate}

\subsection{Case 1: in distribution}\label{case-1-in-distribution}

\begin{figure}
\centering
\subfloat[\label{fig-true}]{\includegraphics[width=0.99\hsize]{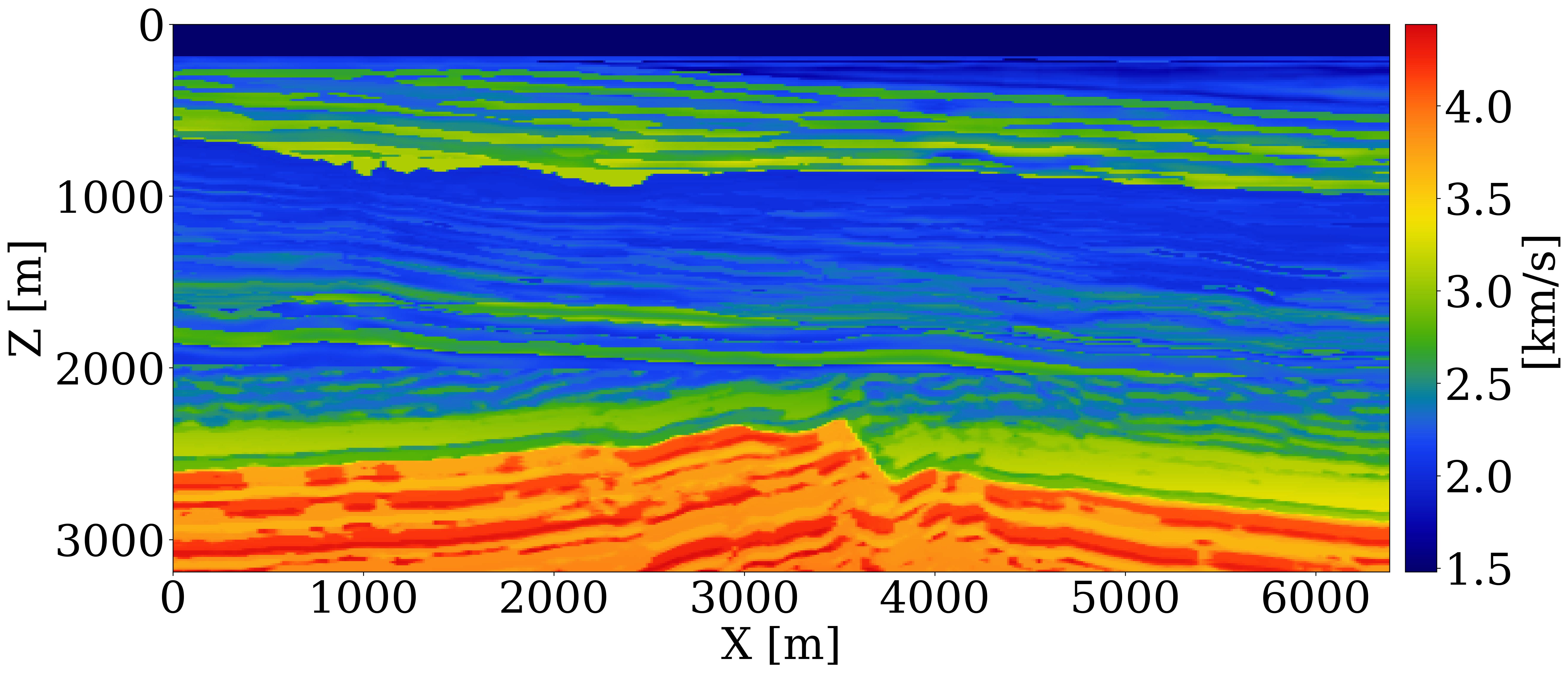}}\\
\subfloat[\label{fig-wise-v-3d}]{\includegraphics[width=0.490\hsize]{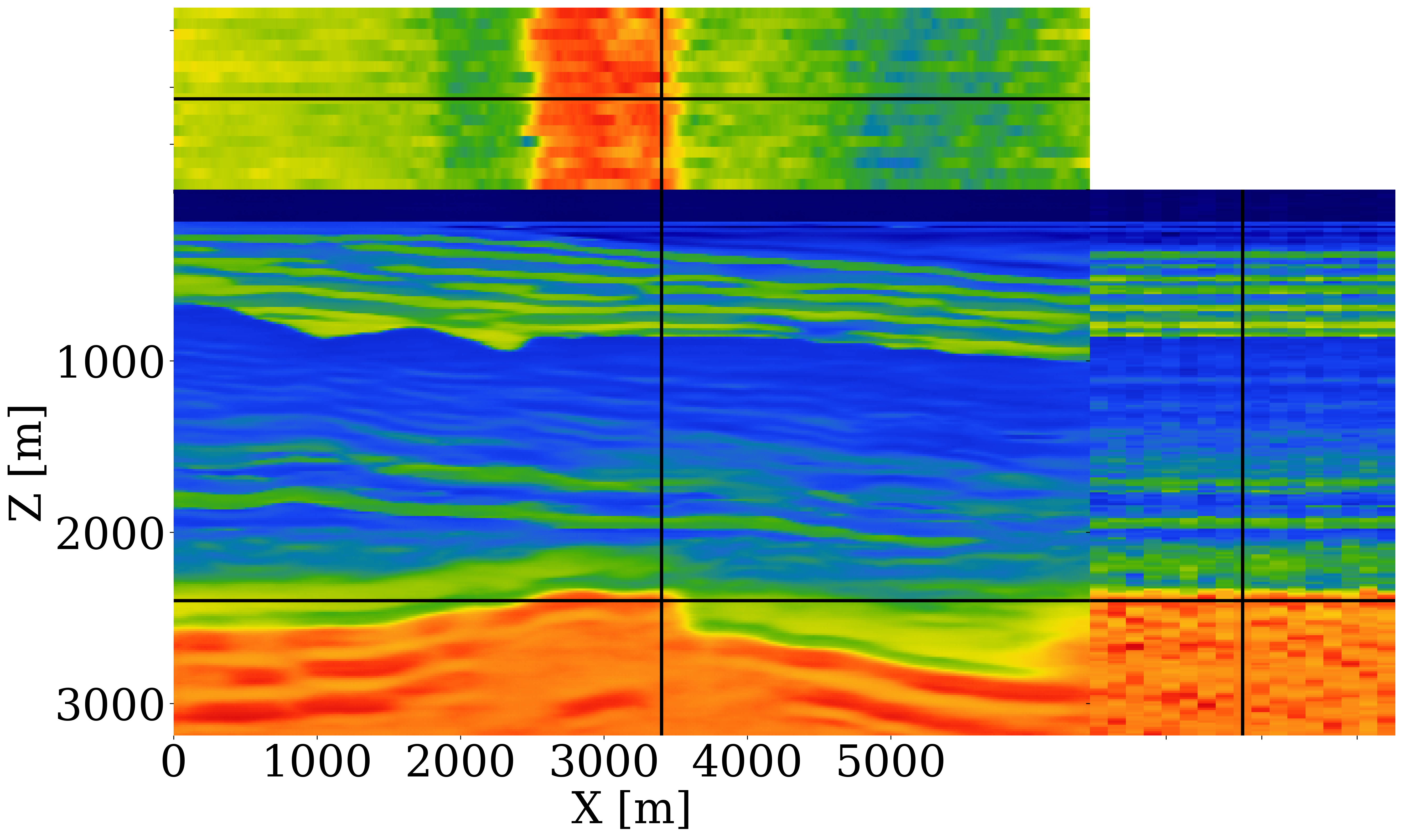}}
\subfloat[\label{fig-wiser-v-3d}]{\includegraphics[width=0.490\hsize]{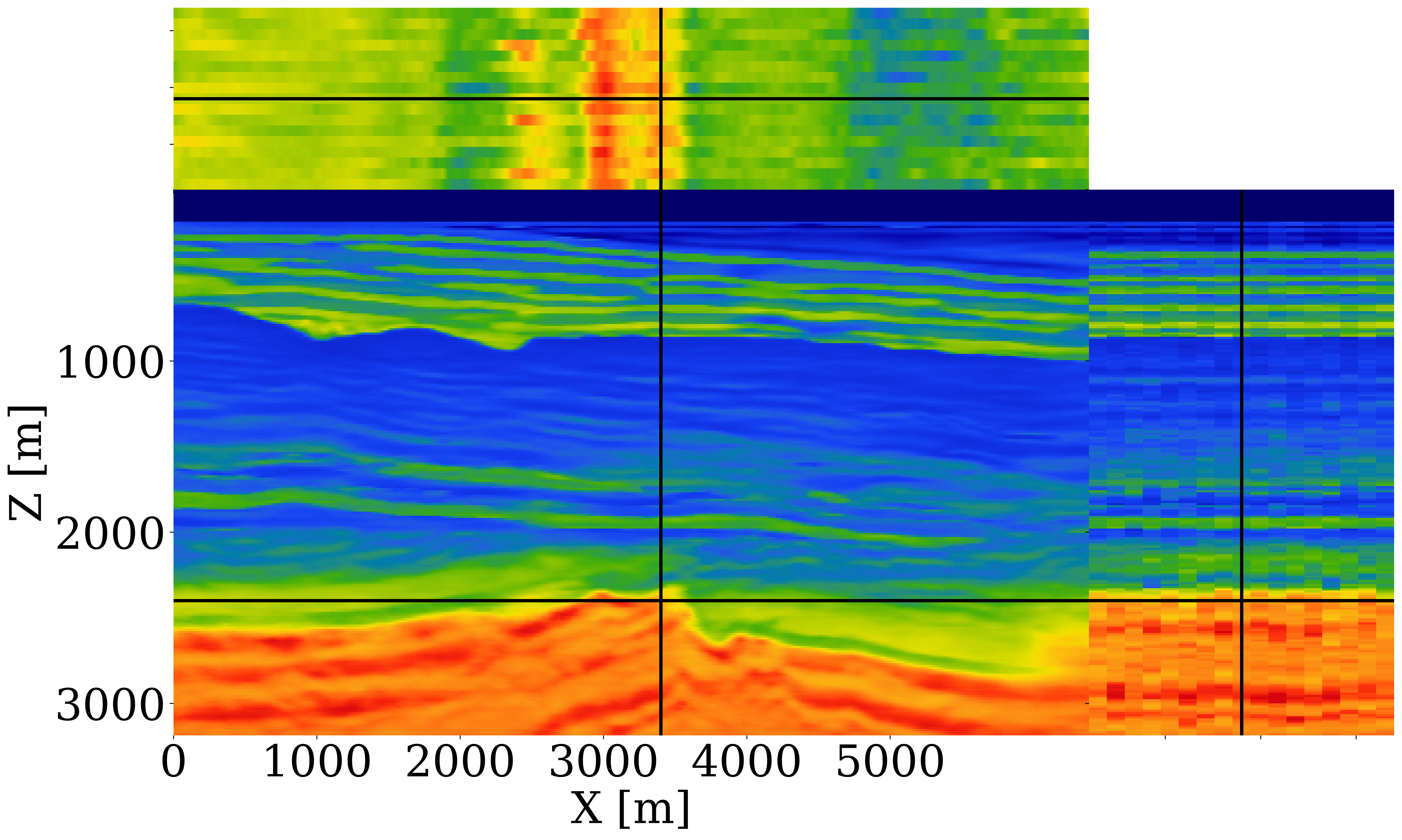}}\\
\subfloat[\label{fig-wise-rtm}]{\includegraphics[width=0.490\hsize]{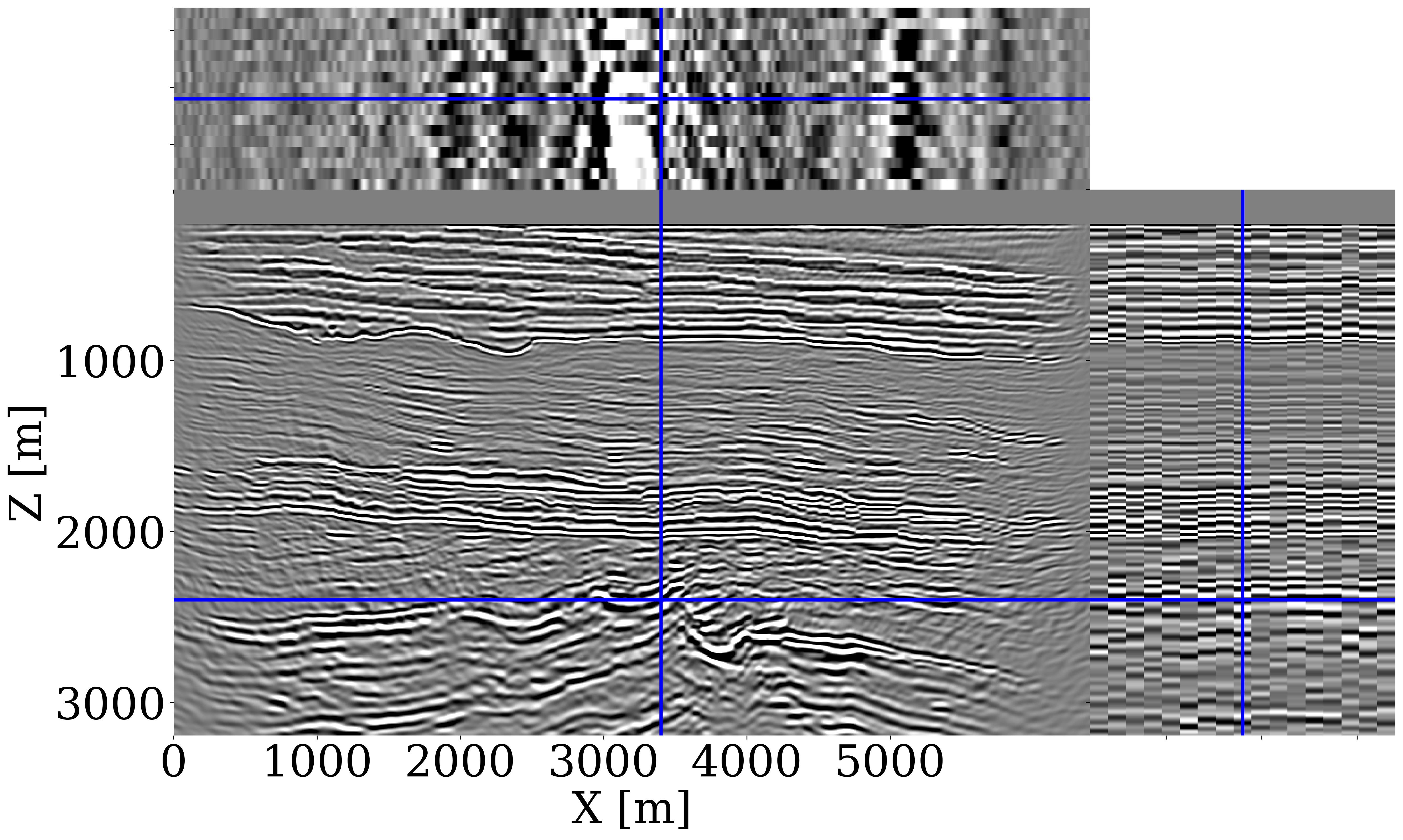}}
\subfloat[\label{fig-wiser-rtm}]{\includegraphics[width=0.490\hsize]{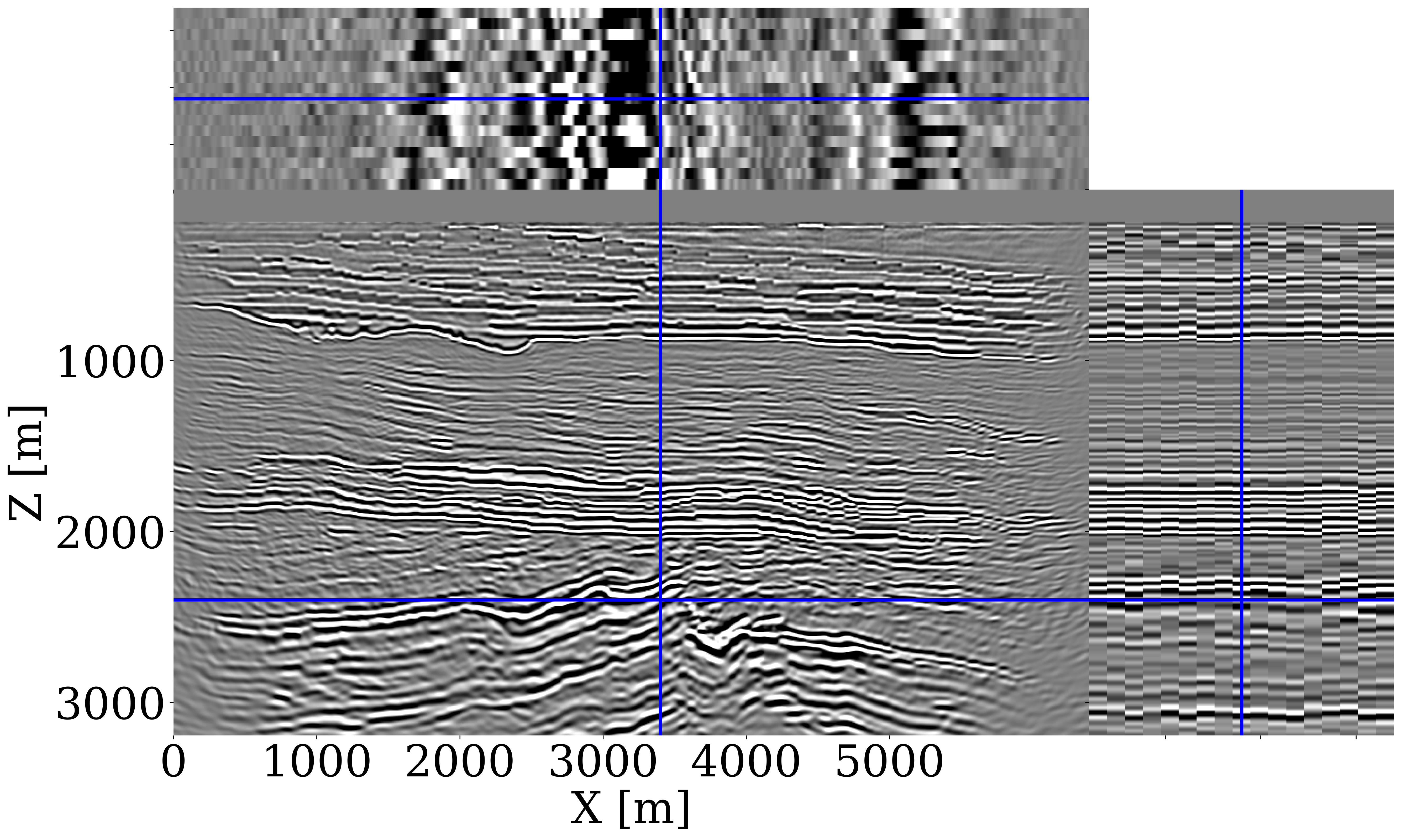}}\\
\subfloat[\label{fig-wise-rtm-zoom}]{\includegraphics[width=0.490\hsize]{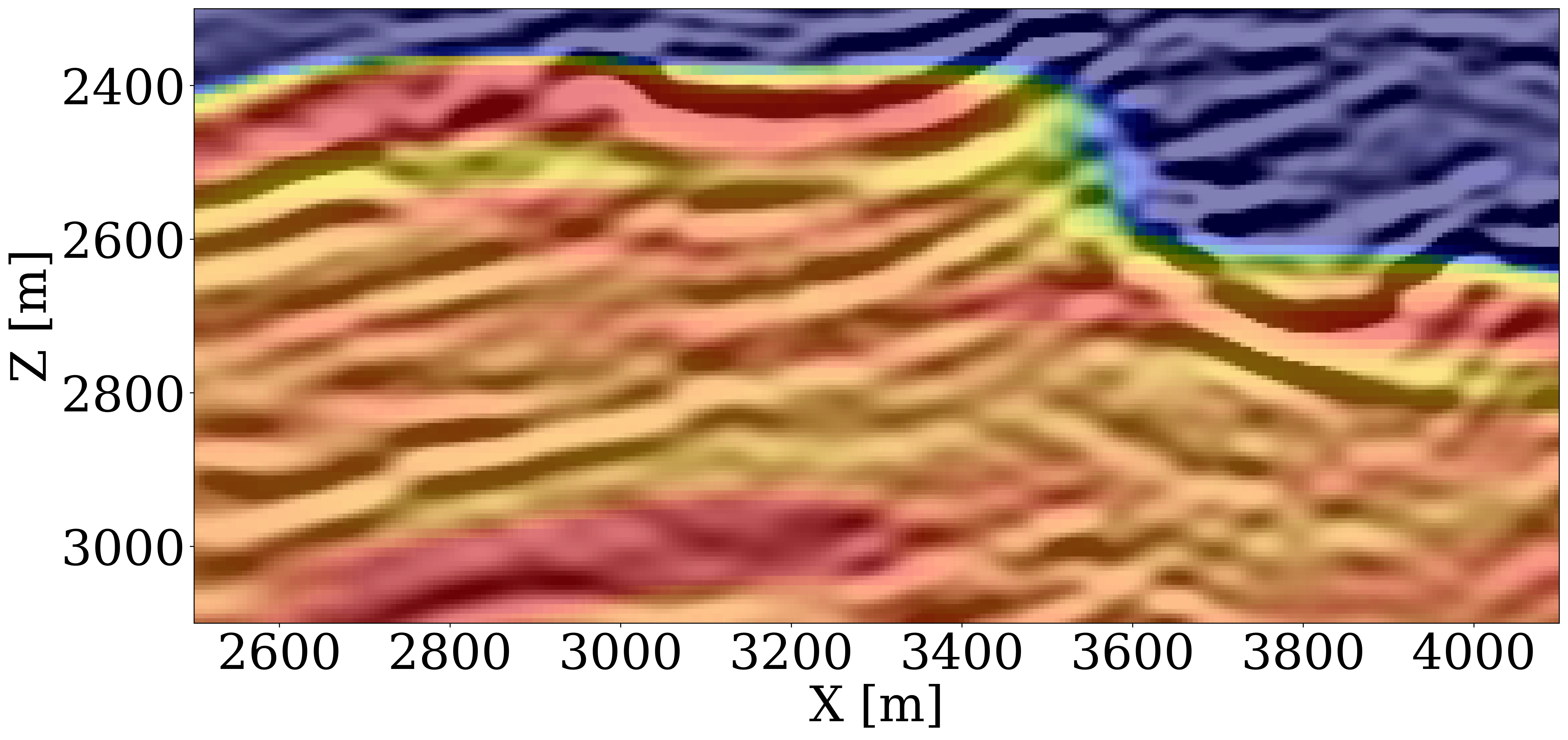}}
\subfloat[\label{fig-wiser-rtm-zoom}]{\includegraphics[width=0.490\hsize]{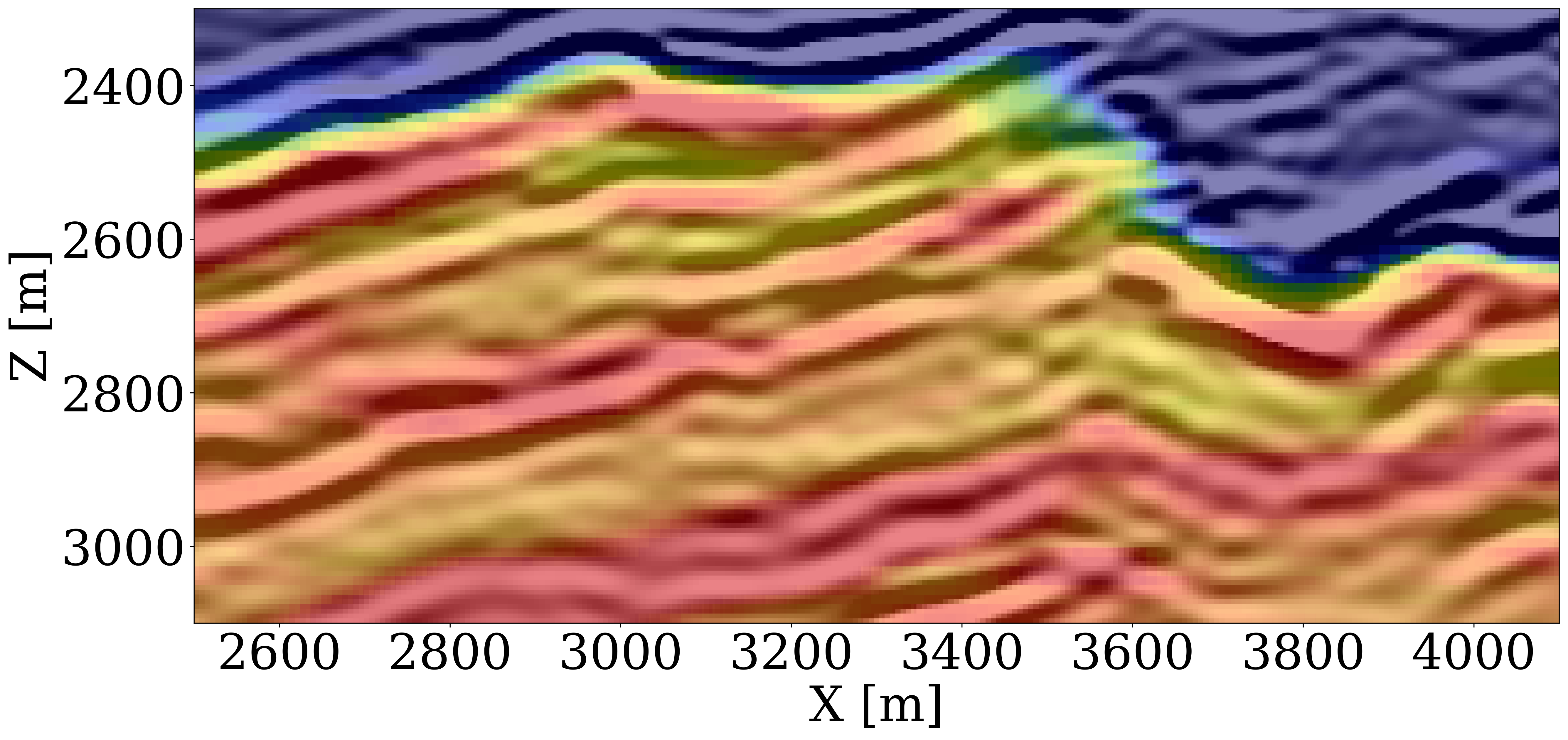}}
\caption{Comparison between WISE and WISER for an in-distribution case.
\emph{(a)} Unseen ground-truth velocity model. \emph{(b)} Estimated
velocity models from WISE. The conditional mean estimate (CM) is shown
in the center. For posterior samples, horizontal traces at
$Z = 2.7\,\mathrm{km}$ and vertical traces at $X = 3.6\,\mathrm{km}$ are
displayed on the top and on the right, respectively. \emph{(d)} Imaged
reflectivity samples from WISE. \emph{(f)} Zoom-in views of \emph{(d)}
overlaying on the CM of WISE. \emph{(c)(e)(g)} are the counterparts from
WISER, showcasing significant improvements.}\label{fig-wise-vs-wiser}
\end{figure}

The ground-truth velocity model is an unseen 2D slice from the Compass
model, shown in Figure~\ref{fig-true}. Following Algorithm~\ref{alg-wiser}, we
initiate WISER by drawing $M=16$ Gaussian noise realizations to create
the initial set of $16$ velocity models, depicted in
Figure~\ref{fig-wise-v-3d}. To minimize computational demands, stochastic
gradients \citep{herrmann2013frugal} are calculated in line 25 of
Algorithm~\ref{alg-wiser}. Each particle's gradient is estimated using only $1$
randomly selected OBN gather from the observed data. We also add box
constraints to the velocity models to restrict their range to $1.48$ to
$4.44\,\mathrm{km/s}$. Following $\mathrm{maxiter}_1=80$ outer
iterations---equivalent to $20$ data passes or $2560$ PDE
solves\footnote{1 PDE solve means solving the wave equation for a single
  source. A gradient requires 2 PDE solves (forward and adjoint).}-----we
obtain the posterior samples from WISER in Figure~\ref{fig-wiser-v-3d}.

\subsubsection{Observations}\label{observations}

The conditional mean estimate (CM) from WISE lacks finer details,
particularly beneath the unconformity at depths below $2.4\,\mathrm{km}$
(in red). This is attributed to the excessive variability in structural
details of the posterior samples, visible on the right panel of
Figure~\ref{fig-wise-v-3d}.

In contrast, WISER generates more consistent and accurate posterior
samples. In Figure~\ref{fig-wiser-v-3d}, the right panel shows that the
uncertainty from WISER is reduced below the unconformity. The upper
panel illustrates that the uncertainty is more focused on the dipping
events at the unconformity, highlighting areas of poor illumination.

\subsubsection{Impact on imaging}\label{impact-on-imaging}

To assess the impact of uncertainty in velocity models on downstream
tasks, we conduct high-frequency imaging using a Ricker wavelet with
central frequency of $30\,\mathrm{Hz}$ and compare the imaged
reflectivities derived from the posterior samples of both WISE and
WISER, shown in Figure~\ref{fig-wise-rtm} and Figure~\ref{fig-wiser-rtm},
respectively.

The imaged reflectivities produced by CM from WISER exhibit superior
continuity and a better correlation with the CM migration-velocity
model, particularly noticeable in Figure~\ref{fig-wiser-rtm-zoom} under the
unconformity. Also, reflectivity samples produced by WISER demonstrate
improved alignment among themselves compared to those produced by WISE.
In addition, notable vertical shifts observed in the imaged
reflectivities from WISE to WISER indicate significant adjustments in
the positioning of subsurface reflectors, underlining the necessity of
the refinement procedure for precisely estimating migration-velocity
models that locate subsurface reflectors more accurately.

\subsection{Case 2: out of
distribution}\label{case-2-out-of-distribution}

\begin{figure}
\centering
\subfloat[\label{fig-perturb-curve}]{\includegraphics[width=0.2\hsize]{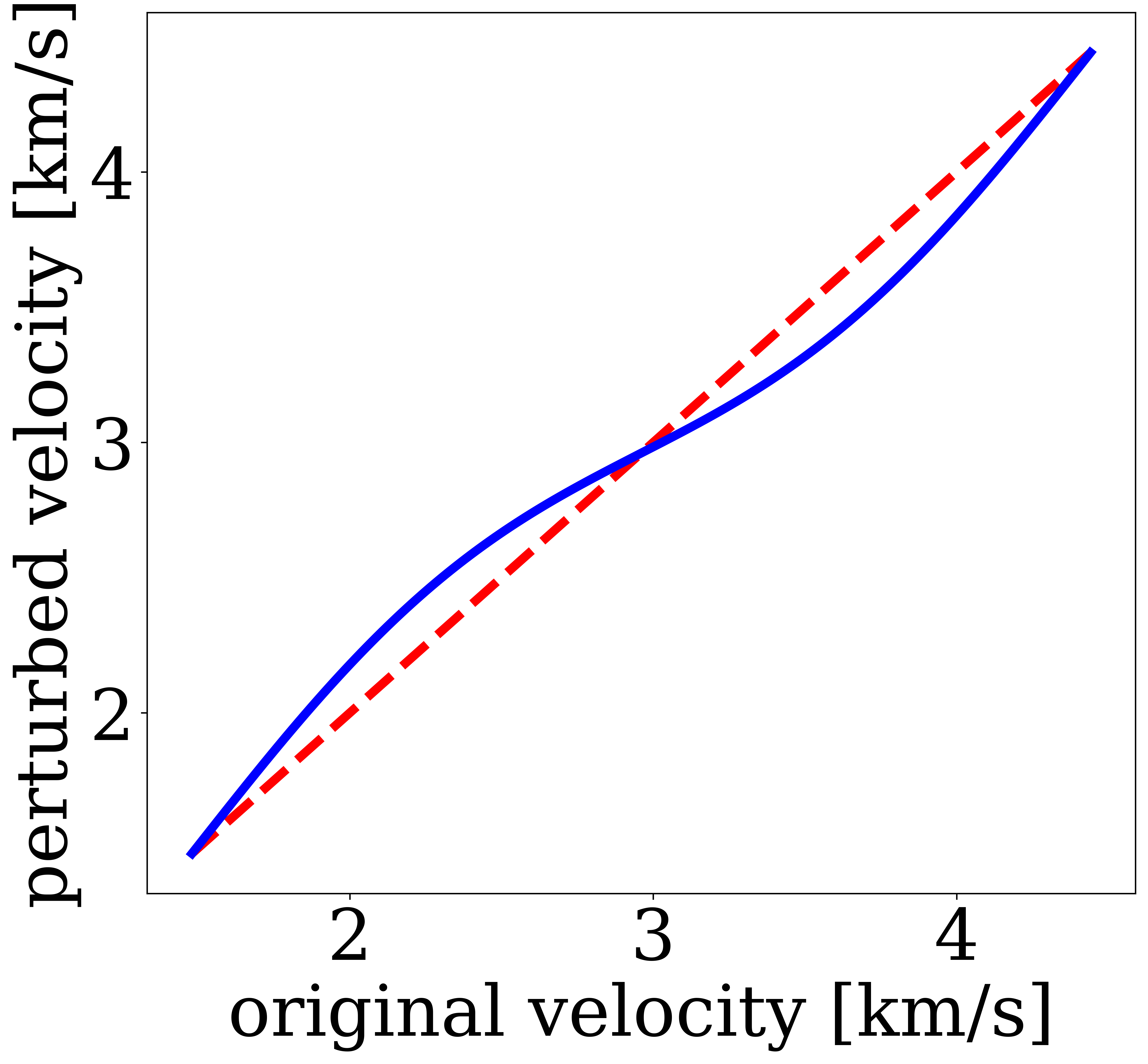}}
\subfloat[\label{fig-hist}]{\includegraphics[width=0.4\hsize]{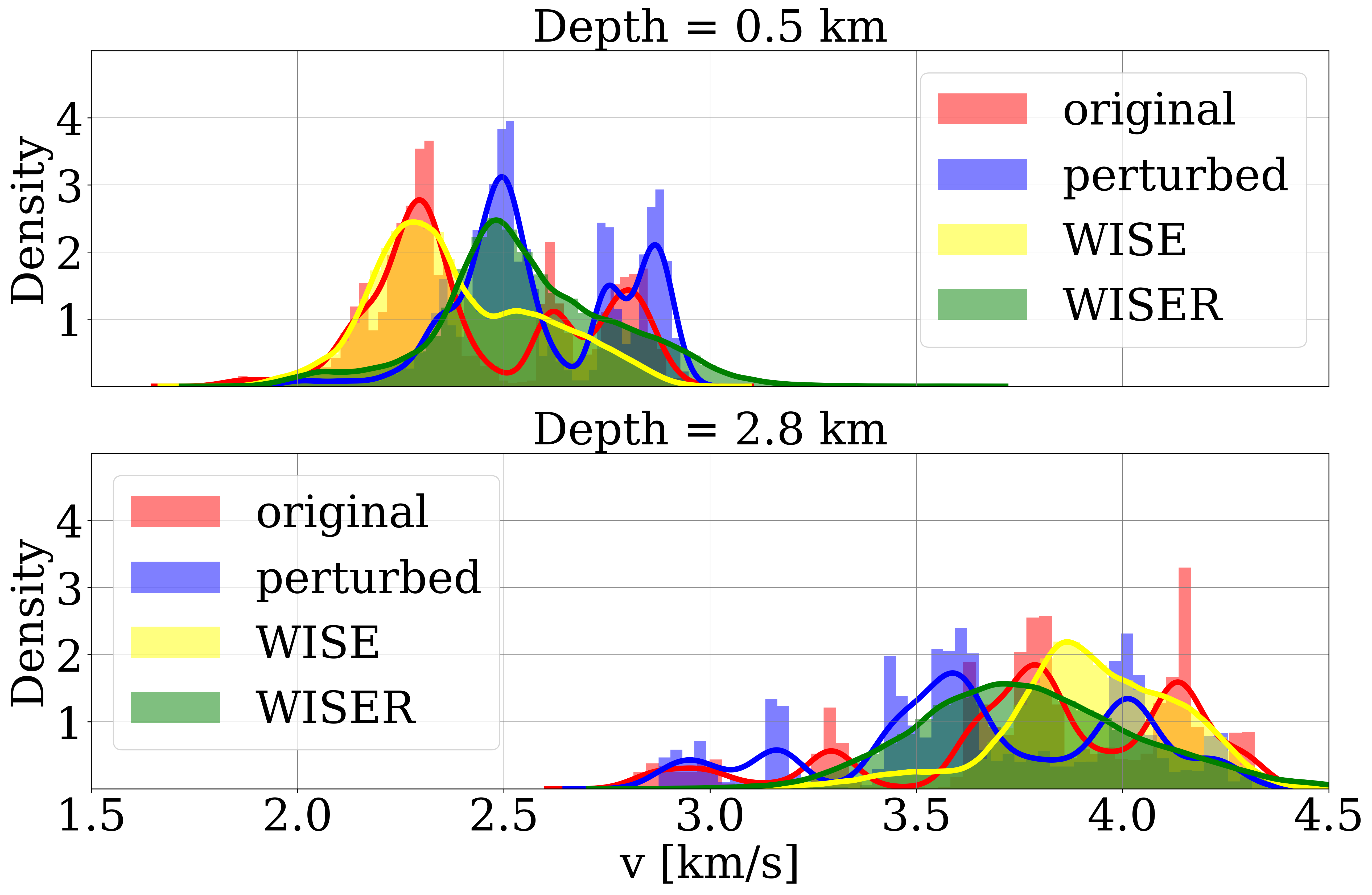}}
\subfloat[\label{fig-v-ood}]{\includegraphics[width=0.39\hsize]{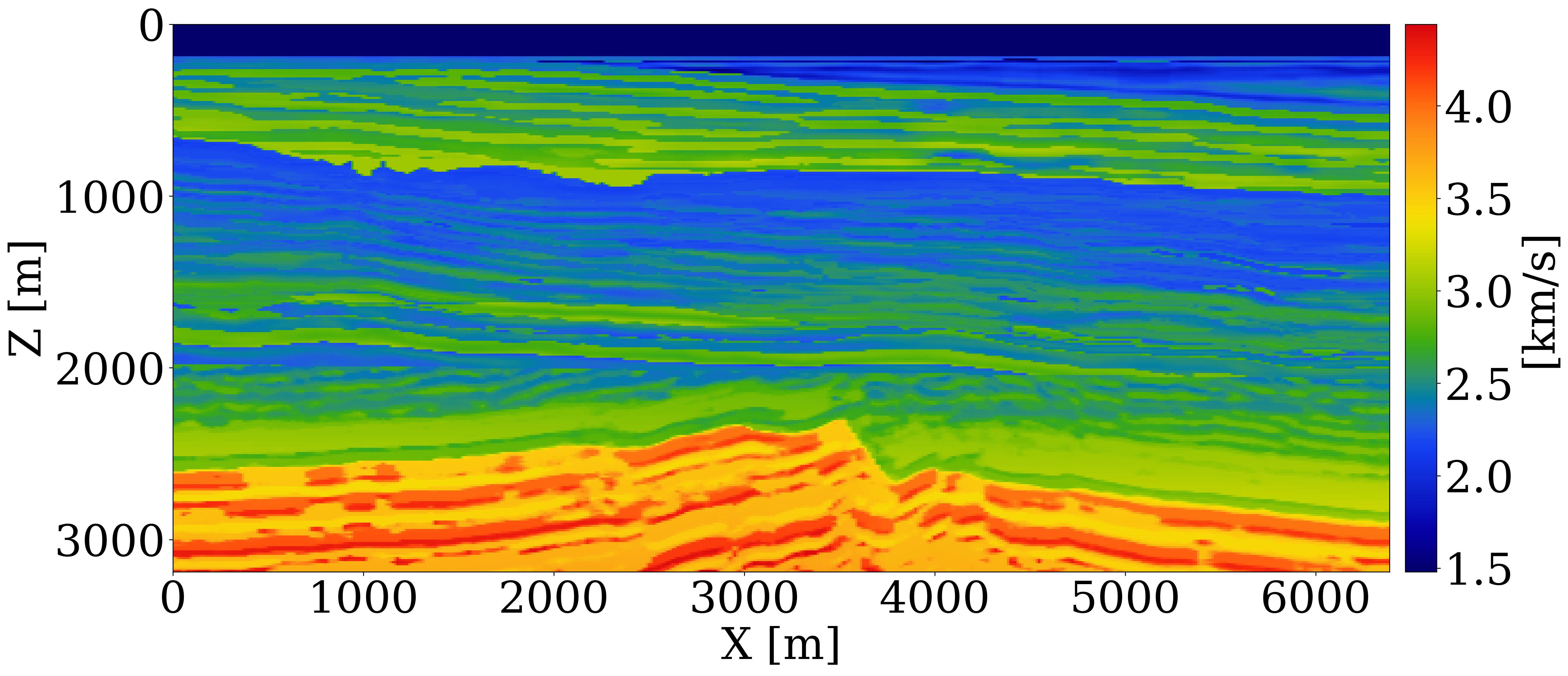}}\\
\subfloat[\label{fig-wise-v-3d-ood}]{\includegraphics[width=0.4\hsize]{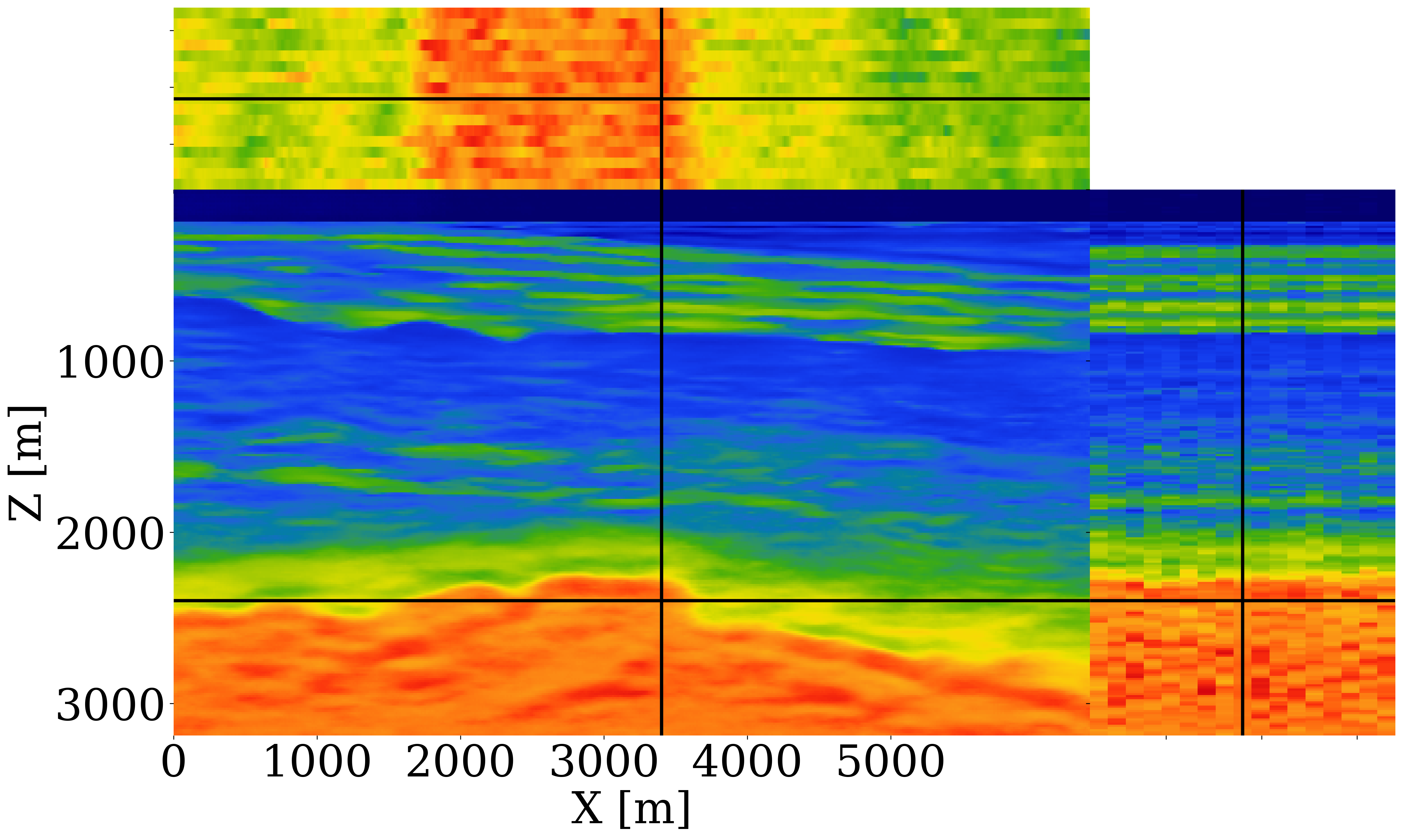}}
\subfloat[\label{fig-wiser-v-3d-ood}]{\includegraphics[width=0.4\hsize]{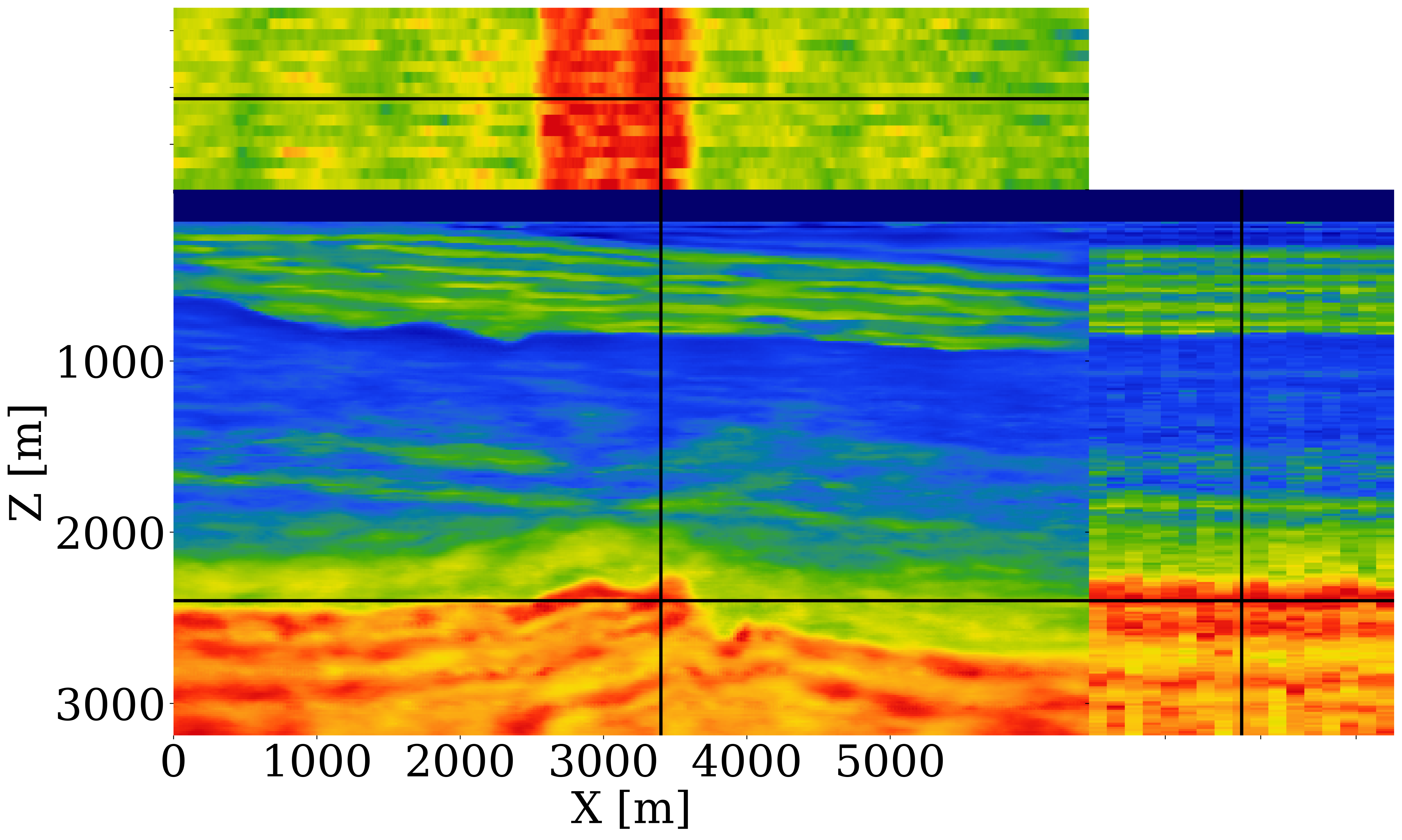}}\\
\subfloat[\label{fig-wise-ood-rtm}]{\includegraphics[width=0.4\hsize]{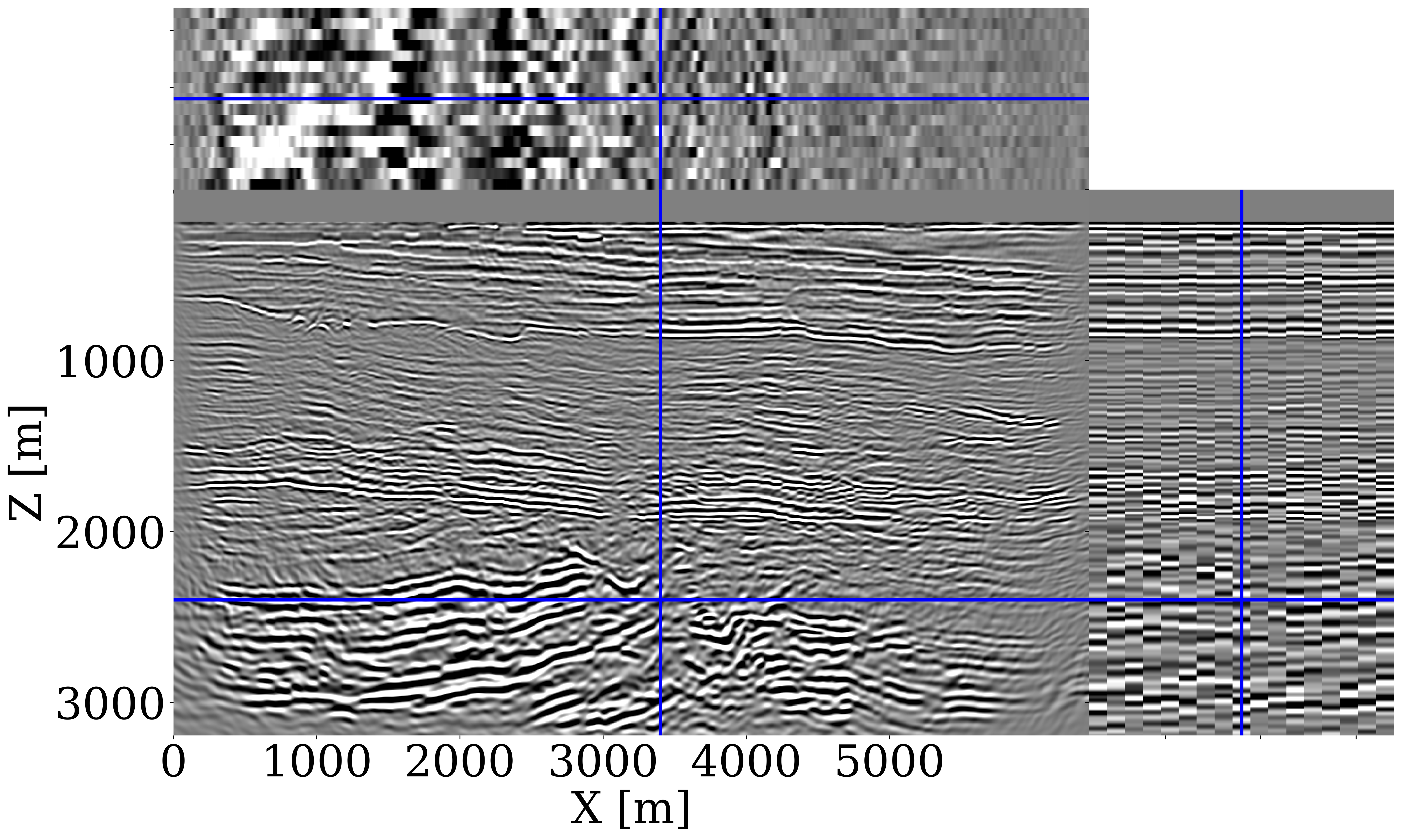}}
\subfloat[\label{fig-wiser-ood-rtm}]{\includegraphics[width=0.4\hsize]{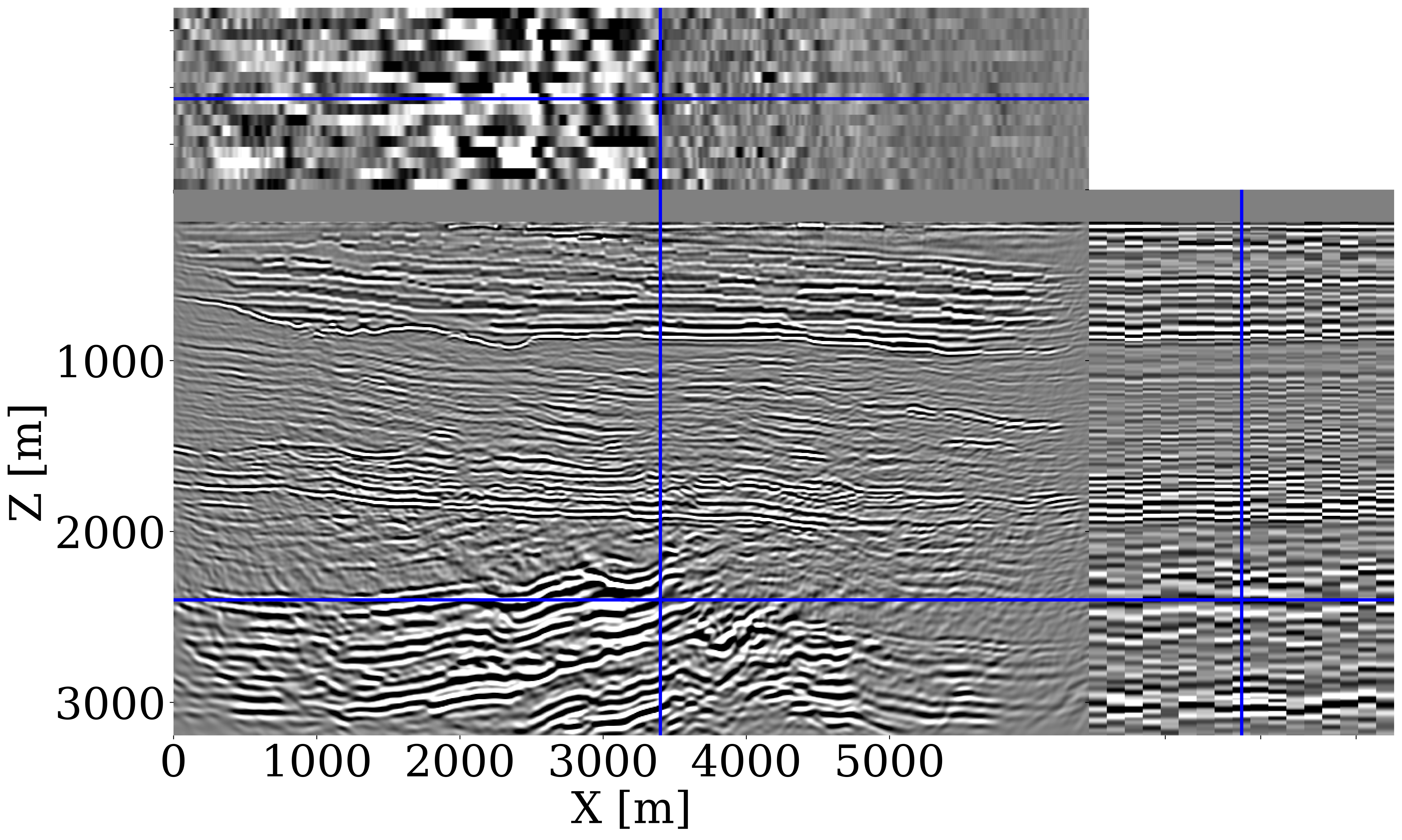}}\\
\subfloat[\label{fig-wise-rtm-zoom-ood}]{\includegraphics[width=0.4\hsize]{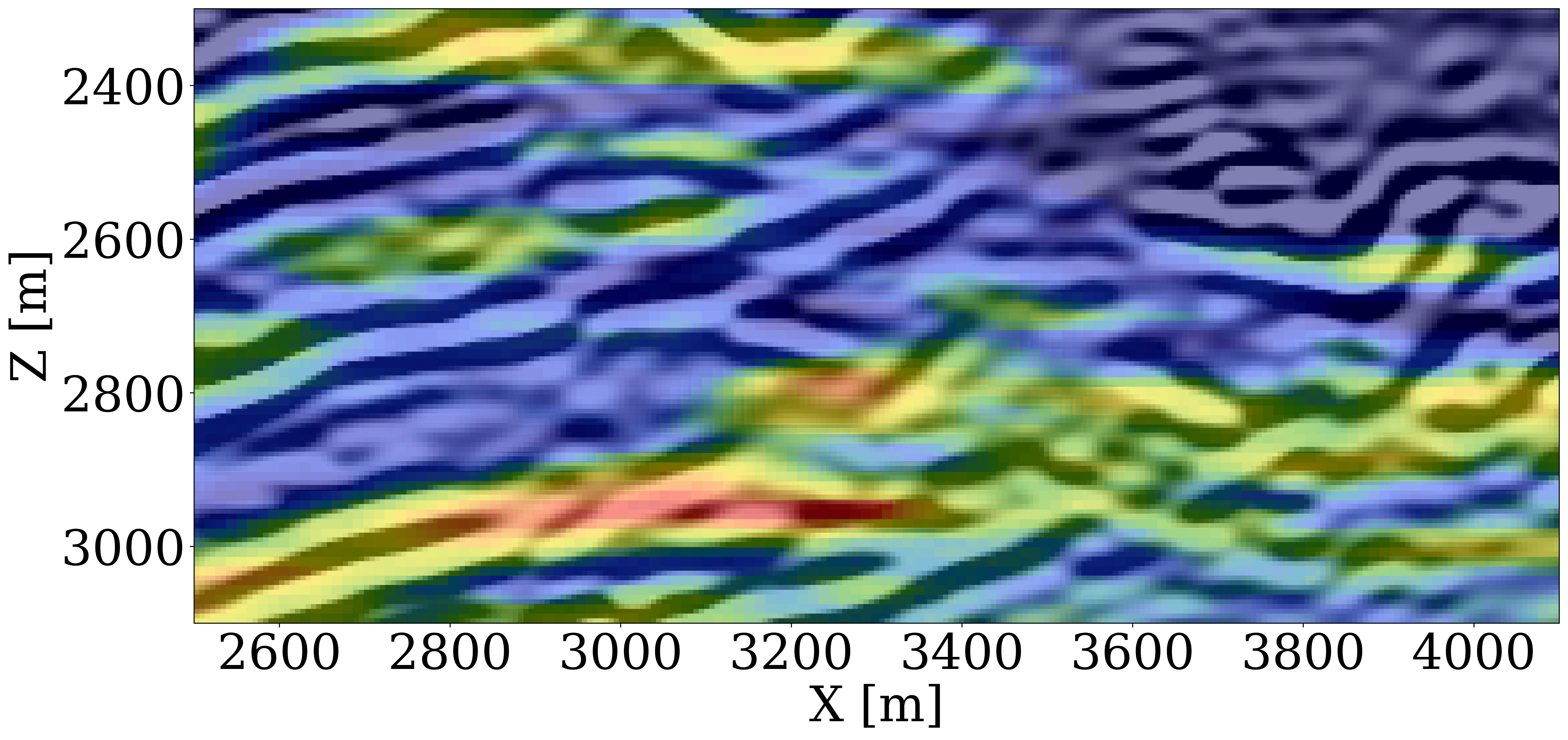}}
\subfloat[\label{fig-wiser-rtm-zoom-ood}]{\includegraphics[width=0.4\hsize]{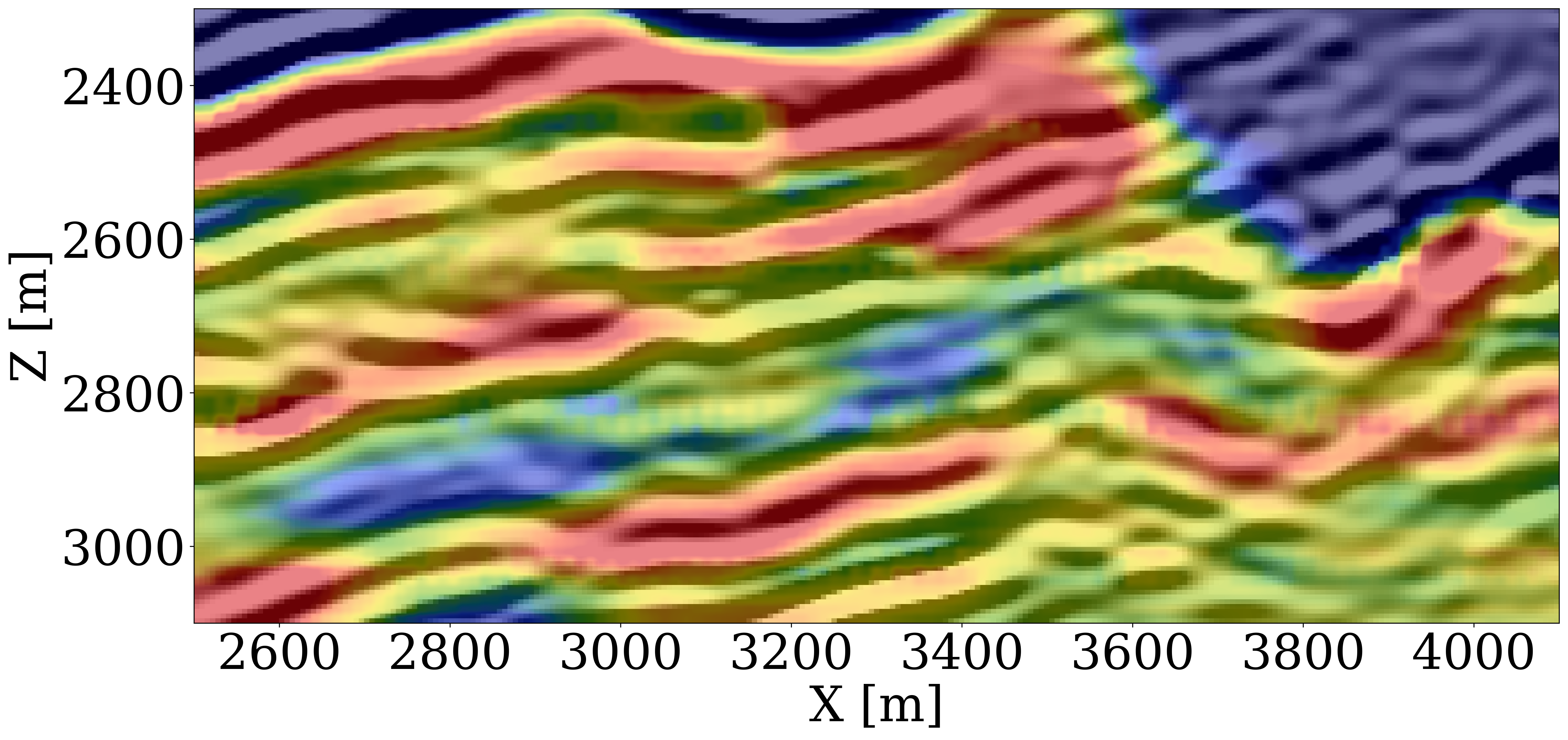}}\\
\subfloat[\label{fig-wise-fwi}]{\includegraphics[width=0.4\hsize]{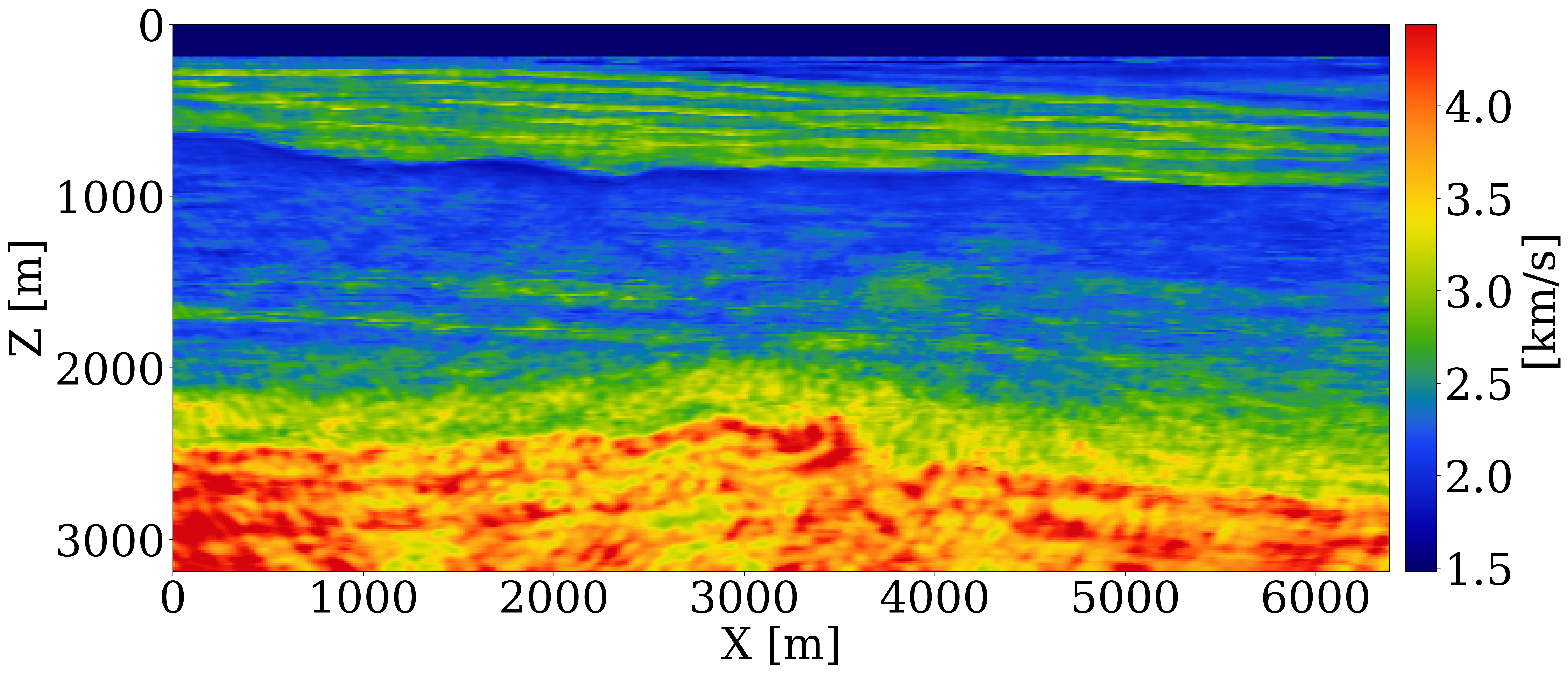}}
\subfloat[\label{fig-wiser-post}]{\includegraphics[width=0.4\hsize]{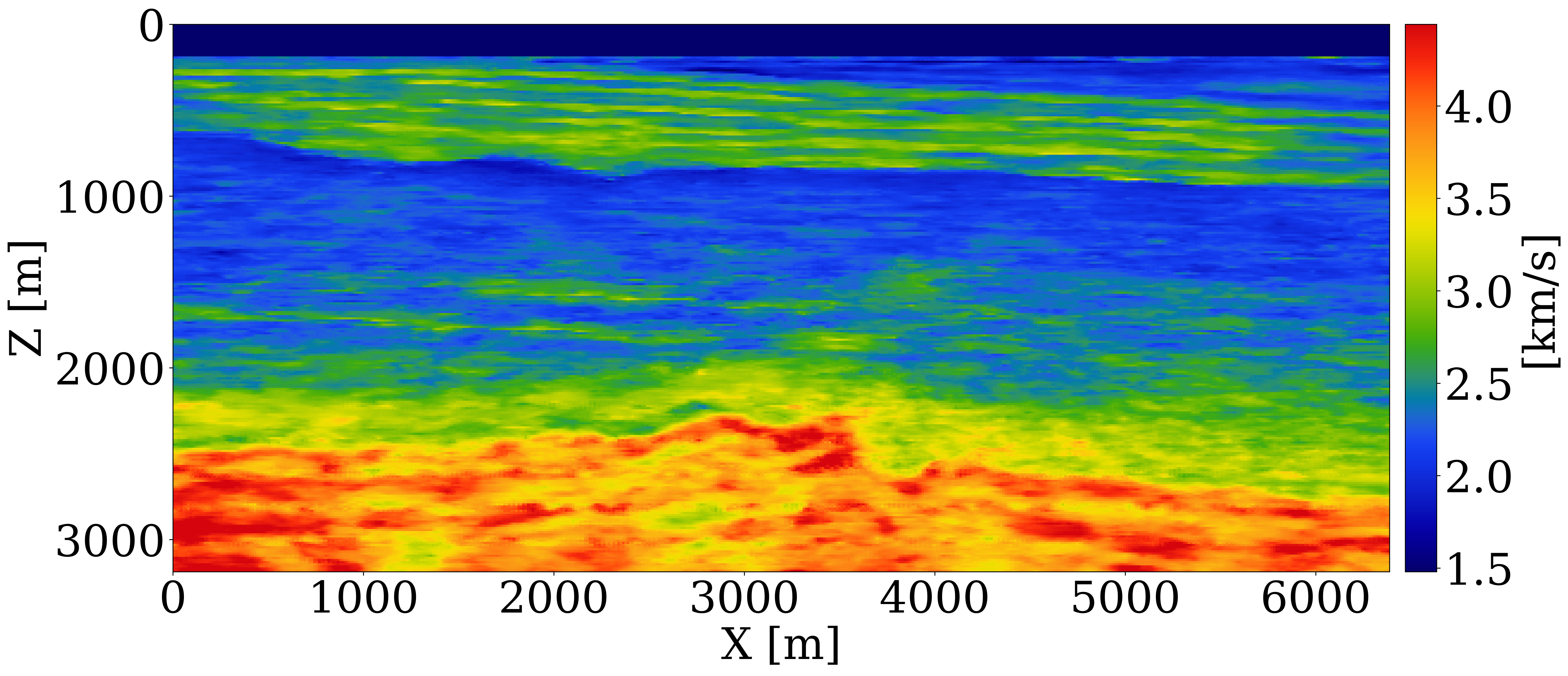}}
\caption{OOD case study. \emph{(a)} Curves for velocity-value perturbations;
\emph{(b)} histograms of values at the depth of $0.5\,\mathrm{km}$ and
$2.8\,\mathrm{km}$ in the original velocity model (Figure~\ref{fig-true}),
perturbed velocity model (Figure~\ref{fig-v-ood}), posterior samples of WISE,
and WISER, shown in red, blue, yellow and green color, respectively.
\emph{(c)---(i)} Comparison between WISE and WISER. The ordering remains
the same as in Figure~\ref{fig-wise-vs-wiser}. \emph{(j)} FWI result starting
with a posterior sample from WISE. \emph{(k)} A posterior sample from
WISER.}\label{fig-wise-vs-wiser-ood}
\end{figure}

To test the robustness and adaptability of WISER when faced with
unexpected variations at inference, we also evaluate WISER's performance
under OOD scenarios. We introduce alterations to the velocity model
depicted in Figure~\ref{fig-true} through an elementwise perturbation shown
in Figure~\ref{fig-perturb-curve}. This manipulation modifies the velocity
values across different depth levels, resulting in a significant shift
in their statistical distribution, illustrated in Figure~\ref{fig-hist}. We
use the perturbed velocity as the unseen ground-truth velocity model in
this case study, shown in Figure~\ref{fig-v-ood}. To further expand the
amortization gap, we modify the encoding of the forward operator by
introducing a higher amplitude of band-limited Gaussian noise (S/N
$0\,\mathrm{dB}$).

These complexities present substantial challenges for WISE, leading to
biased inference results as depicted in Figure~\ref{fig-wise-v-3d-ood}. The
yellow histograms in Figure~\ref{fig-hist} show that the velocity values of
the posterior samples from WISE closely resemble those of the original
velocity model, despite the different distribution of the ground-truth
velocity model. This indicates that WISE tends to incorporate an
inductive bias from the training samples. In WISER, we conduct
$\mathrm{maxiter}_1=160$ outer iterations, using $M=16$ particles and
$1$ OBN per gradient. We also employ the frequency continuation method
\citep{bunks1995multiscale} to compute the gradient in line 25 of
Algorithm~\ref{alg-wiser}, transitioning gradually from low-frequency to
high-frequency data. This results in $40$ datapasses or $5120$ PDE
solves in total.

\subsubsection{Observations}\label{observations-1}

WISER produces more accurate posterior samples shown in
Figure~\ref{fig-wiser-v-3d-ood}. Furthermore, the statistical distribution of
the velocity values in the WISER posterior samples (green histogram in
Figure~\ref{fig-hist}) aligns better with the distribution of the unseen
ground-truth velocity values (blue histogram in Figure~\ref{fig-hist}),
demonstrating WISER's robustness against potential distribution shifts
during inference.

To further showcase WISER's robustness under severe measurement noise,
we compare a posterior sample from WISER (Figure~\ref{fig-wiser-post}) to a
velocity model estimated by FWI (Figure~\ref{fig-wise-fwi}), derived by
minimizing only the data likelihood (the first term in line 25 of
Algorithm~\ref{alg-wiser}), while starting from the same initial model as WISER.
The FWI result is significantly impacted by noise, while the posterior
samples from WISER remain relatively noise-free and capture all
pertinent geological structures.

\subsubsection{Impact on imaging}\label{impact-on-imaging-1}

The imaging results from WISE (Figure~\ref{fig-wise-ood-rtm}) and WISER
(Figure~\ref{fig-wiser-ood-rtm}) reveal noticeable discrepancies in quality.
The CM migration-velocity model from WISE leads to discontinuities in
the imaged reflectivities, particularly at the horizontal layer around
$1.8\,\mathrm{km}$ depth and more so below the unconformity. In
contrast, the CM from WISER significantly improves the continuity of the
imaged reflectivities across the entire seismic section. The imaged
reflectivity samples from WISER not only show better consistency among
themselves but also align more accurately with the estimated CM
migration-velocity model, particularly visible in
Figure~\ref{fig-wiser-rtm-zoom-ood}.

\section{Discussion and conclusions}\label{discussion-and-conclusions}

The primary contribution of WISER is to leverage both genAI and physics
to achieve a semi-amortized VI framework for scalable
($\mathrm{D}\geq 2$) and reliable UQ for FWI even in situations where
local approximations are unsuitable. At its core, WISER harnesses the
strengths of both amortized and non-amortized VI: the amortized
posterior obtained through offline training provides a low-fidelity but
fast mapping, and the physics-based refinements offer reliable and
accurate inference. Both approaches benefit from information
preservation exhibited by CIGs, rendering our inference successful where
conventional FWI fails due to cycle skipping.

Compared to McMC methods that rely on low-dimensional parameterizations,
WISER does not impose intrinsic dimensionality reductions or
simplifications of the forward model. Therefore, WISER is capable of
delivering full-resolution UQ for realistic multi-D FWI problems. CNFs
are primed for large-scale 3D inversion thanks to their invertibility,
which allows for memory-efficient training and inference
\citep{orozco2023invertiblenetworks}.

Compared to non-amortized VI methods, WISER also requires significantly
less computational resources during inference. This is because WISE, as
a precursor of WISER, already provides near-accurate posterior samples,
making the refinement procedure computationally feasible.
\citet{zhang20233} show that non-amortized VI, without access to
realistic prior information, requires $\mathrm{O}(10^6)$ to
$\mathrm{O}(10^8)$ PDE solves, while WISER only needs $\mathrm{O}(10^3)$
PDE solves. Apart from the computational cost reduction, WISER ensures
that the posterior samples realistically resemble Earth models, thanks
to the integration of the conditional prior information from WISE.
Contrary to non-amortized VI, which needs density evaluations to embed
the prior (i.e., $p(\mathbf{x})$ in line 5 of Algorithm~\ref{alg-wiser}) to
produce realistic Earth models, WISER only needs access to samples of
the prior distribution (i.e., $\mathbf{x}^{(i)}$ in line 5).

Opportunities for future research remain. Although case 2 demonstrates
the robustness of WISER concerning OOD issues, these issues could be
fundamentally addressed by diversifying the training set of WISE through
a foundation model \citep{sheng2023seismic}. Also, our OOD case study
has not yet explored scenarios where the likelihood term is more
pathologically misspecified, such as the presence of unremoved shear
wave energy outside the range of the forward operator, which calls for
further investigations. Our approach will also benefit from calibration
of the estimated posterior on which the authors report elsewhere,
including application of WISE(R) in 3D.

In conclusion, this paper sets the stage for deploying genAI models to
facilitate high-dimensional Bayesian inference with computationally
intensive multi-D forward operators. Deep learning and AI have been
criticized for their reliance on realistic training samples, but WISER
alleviates this reliance and still offers computationally feasible and
reliable inference through a blend of offline training and online frugal
physics-based refinements, preparing our approach for large 3D
deployment.

\section{Acknowledgement}\label{acknowledgement}

This research was carried out with the support of Georgia Research Alliance and partners of the ML4Seismic Center.

\section{Related material}\label{related-material}
The scripts to reproduce the experiments are available on the SLIM GitHub page \href{https://github.com/slimgroup/WISER.jl}{\seqsplit{https://github.com/slimgroup/WISER.jl}}.

\section{Declaration of generative AI and AI-assisted technologies in the writing process}\label{gen-ai-declare}

During the preparation of this work, the authors used ChatGPT to refine sentence structures and improve the readability of the manuscript. After using this service, the authors reviewed and edited the content as needed and take full responsibility for the content of the publication.

\bibliography{paper}

\end{document}